\documentclass{WileyMSP-template}
\usepackage{amsmath}
\usepackage{amssymb}
\usepackage{amsfonts}

\usepackage{circledsteps}
\usepackage{xcolor}
\usepackage{hyperref}

\usepackage{ragged2e}

\usepackage{microtype}
\begin{document}

\pagestyle{fancy}
\rhead{\phantom{~~ }}

\title{Topological Engineering of High-Order Exceptional Points through Transformation Optics}

\maketitle

\author{Kaiyuan Wang}
\author{Qi Jie Wang}
\author{Matthew R. Foreman*}
\author{Yu Luo*}

\begin{affiliations}

Kaiyuan Wang, Matthew R. Foreman\\
School of Electrical and Electronic Engineering, Nanyang Technological University, 50 Nanyang Avenue, Singapore 639798 \\
Institute for Digital Molecular Analytics and Science, 59 Nanyang Drive, Singapore 636921 \\
Email Address: matthew.foreman@ntu.edu.sg\\

Qi Jie Wang\\
School of Electrical and Electronic Engineering, Nanyang Technological University, 50 Nanyang Avenue, Singapore 639798 \\
Centre for Disruptive Photonic Technologies, School of Physical and Mathematical Sciences,
Nanyang Technological University, Singapore, Singapore\\

Yu Luo\\
National Key Laboratory of Microwave Photonics, Nanjing University of Aeronautics and Astronautics, Nanjing 211106, China \\
Email Address: yu.luo@nuaa.edu.cn\\

\end{affiliations}


% Abstract should be written in the present tense and impersonal style (i.e., avoid we), and be at most 200 words long
\begin{abstract}
\justifying % Add this command to achieve alignment
Exceptional points (EPs) in non-Hermitian photonic systems have attracted considerable research interest due to their singular eigenvalue topology and associated anomalous physical phenomena. These properties enable diverse applications ranging from enhanced quantum metrology to chiral light-matter interactions. Practical implementation of high order EPs in optical platforms however remains fundamentally challenging, requiring precise multi-parameter control that often exceeds conventional design capabilities. This work presents a novel framework for engineering high order EPs through transformation optics (TO) principles, establishing a direct correspondence between mathematical singularities and physically controllable parameters. Our TO-based paradigm addresses critical limitations in conventional Hamiltonian approaches, where abstract parameter spaces lack explicit connections to experimentally accessible degrees of freedom, while simultaneously providing full mode solutions. In contrast to prevailing parity-time-symmetric architectures, our methodology eliminates symmetry constraints in EP design, significantly expanding the possibilities in non-Hermitian photonic engineering. The proposed technique enables  control over EP formation and evolution in nanophotonic systems, offering new pathways for developing topological optical devices with enhanced functionality and robustness.
\end{abstract}
\twocolumn
\section{Introduction\label{sec:intro}}
\justifying

An exceptional point (EP) is a branch point singularity that can occur in the spectrum of non-Hermitian systems. An $N$-th order EP (denoted henceforth as EP$N$) corresponds to a point in the system's parameter space at which $N$ eigenvalues and the associated eigenstates of the non-Hermitian Hamiltonian coalesce \cite{Miri 2019, Heiss 2012}. EPs have been extensively studied in recent years as they can give rise to unique physical phenomena, such as unidirectional invisibility \cite{Feng 2013, Regensburger 2012}, enhanced sensitivity \cite{Chen 2017, Loughlin 2024}, and non-reciprocal light propagation \cite{Peng 2014}. Advances in this field have also revealed a variety of complex EP geometries in parameter space, including exceptional rings \cite{Zhen 2015, Cerjan 2019}, arcs \cite{Shen 2018, Zhou 2018}, surfaces \cite{Zhong 2019}, and junctions \cite{Bergholtz 2019}, which can exhibit diverse features like fractional topological charges \cite{Song 2019, Cerjan 2018} and an  anisotropic response to external perturbations \cite{Pan 2019, Bergholtz 2021}. These properties offer significant opportunities to tailor the response of physical systems.

Nano- and micro-optical systems, including microcavities, waveguides, gratings, and nanoplasmonic structures \cite{Yang 2020, El-Ganainy 2018}, have proven to be rich and flexible platforms for realizing, studying, and engineering EPs. For example, the observation of EPs in microcavity systems has enabled enhanced optical sensing capabilities \cite{Chen 2017, Hodaei 2017}, while EPs in waveguides have facilitated the development of novel light manipulation techniques \cite{Doppler 2016, Mandal 2021}.  Current methods to design EPs in optical systems however primarily focus on the system Hamiltonian. Whilst effective in theoretical studies, the Hamiltonian approach faces significant challenges when applied to EP design in realistic physical systems. For instance, Hamiltonians considered in EP design are often limited to idealised or simplified models which do not accurately represent the complexity or capture intricate interactions present in practical systems, especially at the nanoscale. Such issues are further compounded in the design of high order exceptional points, which requires precise adjustment of $2N-2$ real parameters in general,  which is challenging in practice \cite{ Mandal 2021, Ding 2022}. Although specific symmetries, such as pseudo-Hermitian, or chiral symmetry, can help reduce the difficulty of finding EPs \cite{Sayyad2022, Montag2024}, generalizing these symmetries to the generation of higher-order EPs remains an open problem \cite{Hassan 2015, Liu 2012}.

In this work, we propose an approach to design high order EPs which leverages transformation optics (TO) and is capable of alleviating limitations of the Hamiltonian based approach. TO is a powerful and versatile method that allows manipulation of electromagnetic fields through engineering material properties \cite{Chen 2010}. The flexibility of TO allows us to not only model complex geometries but also to design EPs with specific mode distributions. Our approach moreover does not rely on PT symmetry, thereby enabling the design of non-PT symmetric EPs, which have garnered significant attention due to their broader range of unique properties \cite{El-Ganainy 2018, Makris 2008, McCall 2018, Sun 2017}.

\section{Design principles and results \label{sec:design}}

Transformation optics exploits the invariance of Maxwell's equations under coordinate transformations, such that known electromagnetic solutions for simple geometries can be mapped to more complex structures \cite{Kundtz 2010,Pendry 2006}. By first deriving the canonical resonance condition for modes in a planar multilayered structure, we can  conveniently tailor the physical parameters of the system to generate high order EPs, before the solution is finally mapped to a more complicated geometry. In this work we focus on design of EPs in nanoplasmonic systems. Noting that in a multilayer geometry, each interface can potentially support a surface plasmon polariton (SPP) mode, an $N$th order EP can thus in principle be designed for a system with $N$ interfaces ($N+1$ layers) \cite{Mandal 2021}. Existence of an $N$th order EP, however, requires that the resonance equation be expressible as an $N$th order polynomial in one of the frequency dependent system parameters (typically electric permittivity), and that this polynomial possesses $N$ repeated roots. These constraints help guide system design and allow us to determine suitable material and geometric parameters to realise high order EPs.

\begin{figure*}[t]
	\begin{center}
	\includegraphics[width=0.8\textwidth]{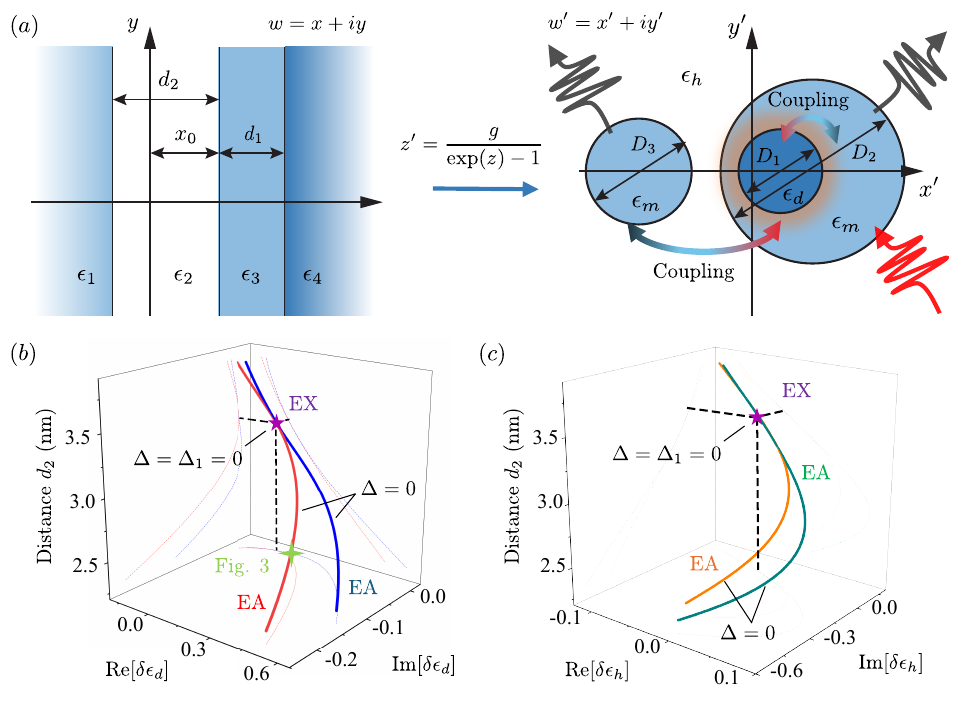}
	\caption{\textbf{TO approach to designing exceptional points.} (a) Schematic diagram of planar geometry considered before (left) and plasmonic nanowire geometry consider after (right) transformation. (b) Exceptional arcs (EA) of exceptional points (EPs), defined by $\Delta = 0$ (see main text), traced in physical parameter space upon variation of the electric permittivity of gain material and interface separation $d_2$. The intersection of the two EAs corresponds to an exceptional nexus (EX), a point at which a higher order EP exists.  (c) As (b) albeit for variations in host material. \label{fig:schematic}}
	\end{center}
\end{figure*}

For concreteness, we illustrate the approach through construction and analysis of a third-order EP in a coupled core-shell/monomer nanowire system as depicted in Figure~\ref{fig:schematic}(a). We therefore begin by  considering a general two-dimensional four layer system, with electric permittivities $
\epsilon_j$ for integer $j\in [1,4]$ and interface separations $d_1$ and $d_2$ as shown in Figure~\ref{fig:schematic}(a). In the quasi-static limit, we can express supported modes in terms of their scalar potential distribution, from which we can derive the corresponding electric and displacement fields, $\mathbf{E}$ and $\mathbf{D}$ respectively (see the Supplementary Material for further details and alternative derivations). Application of standard boundary conditions for the tangential (normal) components of the electric (displacement) field  at each interface yields a set of linear equations relating the tangential electric field component at each interface, specifically
\begin{equation}
	\mathbb{M}^{(3)}\mathbf{E}^{(3)} = \mathbf{0}
	\label{eq:ME_3nd_order}
\end{equation}
where
\begin{align}
   \mathbb{M}^{(3)}= \left[\begin{array}{ccc}
		\mathcal{M}(d_2,\epsilon_2,\epsilon_1)  & -\epsilon_2 &0\\
		0 &-\epsilon_3 &\mathcal{M}(d_1,\epsilon_3,\epsilon_4)\\
		\mathcal{M}(d_2,\epsilon_1,\epsilon_2) & 0 &
		\mathcal{M}(d_1,\epsilon_4,\epsilon_3)
	\end{array}  \right],
\end{align}
\begin{align}
    \mathbf{E}^{(3)} =
	\left[\begin{array}{c} E_{y}(x_0-d_2)\\
		E_y(x_0)\\
		E_y(x_0+d_1)\end{array}\right],
\end{align}
$\mathcal{M}(d,\epsilon_a,\epsilon_b) =\epsilon_a \mbox{cosh}(|k|d) + \epsilon_b \mbox{sinh}(|k|d)$ and $k$ is the {tangential wavenumber along the interface}. Resonances correspond to non-trival solutions of Eq.~\eqref{eq:ME_3nd_order} which occur when
\begin{align}
	\mbox{det}[\mathbb{M}^{(3)}]&= \epsilon_2 \mathcal{M}(d_1,\epsilon_3,\epsilon_4)\mathcal{M}(d_2,\epsilon_1,\epsilon_2) \nonumber\\ &\quad+ \epsilon_3 \mathcal{M}(d_1,\epsilon_4,\epsilon_3) \mathcal{M}(d_2,\epsilon_2,\epsilon_2)  = 0. \label{eq:3rd_resonance_condition}
\end{align}
Eq.~\eqref{eq:3rd_resonance_condition} in its most general form is quadratic in the permittivity of any given layer and therefore it is not readily apparent how to design EPs of order higher than 2. By asserting, however, that two layers have identical permittivity (whilst keeping the number of interfaces fixed), Eq.~\eqref{eq:3rd_resonance_condition} can be expressed as a cubic equation. Specifically, if we assume $\epsilon_1 = \epsilon_3 = \epsilon_m$ (which we take as a lossy metal in what follows), $\epsilon_2 = \epsilon_h$ (a `host' dielectric material) and $\epsilon_4 = \epsilon_d$ (dielectric),  Eq.~\eqref{eq:3rd_resonance_condition} can be written as
\begin{equation}
A\epsilon_m^3 + B\epsilon_m^2 + C\epsilon_m + D = 0, \label{eq:cubic}
\end{equation}
where the coefficients are defined in the Supplementary Material.  Notably, these material choices do not correspond to a PT symmetric system.  Furthermore, note that Eq.~\eqref{eq:cubic} can be considered as the characteristic equation of some eigenvalue problem $|\mathbb{A} - \epsilon_m \mathbb{I}|=0$, such that we will refer to values of $\epsilon_m$ satisfying Eq.~\eqref{eq:cubic} (denoted $\epsilon_{m,r}$ for $r=1,2,3$) as eigenvalues. The corresponding frequencies $\omega_{r}$ for which $\epsilon_m(\omega_{r}) = \epsilon_{m,r}$ shall be termed resonant eigenfrequencies. To generate a third order degeneracy we apply the well known criteria for a cubic equation to have three repeated roots. Specifically, letting the discriminant $\Delta = (4\Delta_1^3 - \Delta_2^2)/(27A^2)$, where $\Delta_1 = B^2 - 3AC$ and $\Delta_2 = 2B^3 - 9ABC+27A^2D$ \cite{cubicquartic}, we require $\Delta = \Delta_1 = 0$. For our example, the repeated root is then given by
\begin{align}
	\epsilon_m = - \frac{B}{3A} &= -\frac{1}{3}\big(\epsilon_h[1+\coth (kd_1)]\coth(kd_2) \nonumber \\
 &\quad\quad\quad\quad+ \epsilon_d \coth(kd_1)\big). \label{eq:EP3constraints}
\end{align}
Noting that the metal is lossy and that $k$, $d_1$ and $d_2$ are real positive numbers, Eq.~\eqref{eq:EP3constraints} implies that either $\epsilon_h$ or $\epsilon_d$ (or both) must have an imaginary part of opposite sign to $\epsilon_m$, i.e. must possess gain. Henceforth, we consider the case $\mbox{Im}[\epsilon_h] = 0$ and that $\epsilon_d$ describes a material with gain, as depicted in Figure~\ref{fig:schematic}(a).
 According to the principles of TO, the same conditions also govern the mode spectrum in the coupled nanowire geometry shown on the right of Figure~\ref{fig:schematic}(a), which can be generated through application of the exponential conformal mapping \cite{Aubry 2011}:
\begin{equation}
w' = \frac{g}{\exp(w) - 1} \label{eq:conformal_transform}
\end{equation}
where $w^{(\prime)} = x^{(\prime)}+iy^{(\prime)}$, $(x,y)$ are the untransformed spatial coordinates and primes denote transformed quantities. The parameter $g$ is a scaling constant that controls the diameter of the nanowires ($D_1$, $D_2$, $D_3$ defined in the Supplement) in the transformed geometry. We assume that $g$ is chosen such that the dimensions of the nanowire system are smaller than the optical wavelength, whereby the quasi-static model holds. Within this limit, the change in permeability after transformation can also be ignored \cite{Pendry 2013}. Finally, we note, that the periodicity of the transformation implies $k = n$ ($n\in \mathbb{Z})$, which  physically corresponds to the angular momentum of the corresponding mode.

 \begin{figure*}[t!]
 	\begin{center}
 	\includegraphics[width=\textwidth]{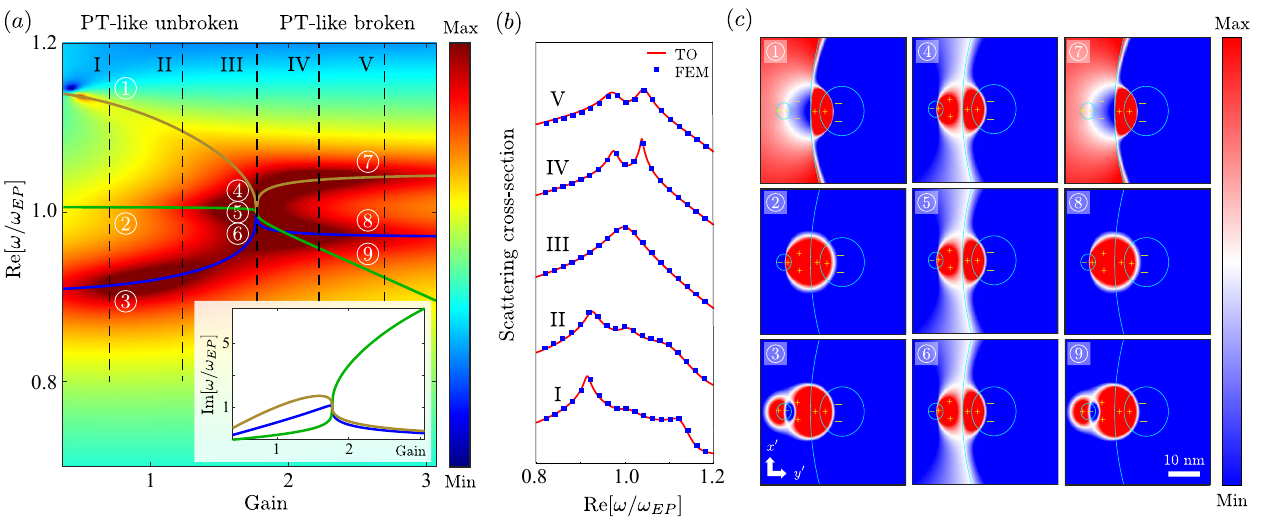}
 	\caption{{\textbf{EP3 in core-shell/monomer coupled nanowire system.}} (a) Color density plot of the logarithm of the scattering cross-section of the coupled core-shell and monomer structures as a function of gain and frequency. Solid lines represent the real parts of the three eigenfrequencies (corresponding imaginary parts are shown in the inset). (b) scattering cross-sections corresponding to I-V in (a) as found using our TO approach (red lines) and finite element simulations (blue markers). (c) Potential distributions for eigenmodes at points labelled \Circled{1}-\Circled{9} in (a). The $+$ and $-$ markers denote a positive and negative potential respectively. Cyan lines demark the nanowire interfaces (only a small part of core-shell nanowire in vicinity of monomer is shown to enhance visibility). Panels \Circled{4}-\Circled{6} correspond to the EP3.  }
 	\label{fig:EP3}
 	\end{center}
 \end{figure*}

 For material and geometric parameters satisfying $\Delta=\Delta_1=0$ and Eq.~\eqref{eq:EP3constraints}, the solutions of Eq.~\eqref{eq:3rd_resonance_condition} are threefold degenerate. Accordingly, these equations describe necessary parameteric constraints which must be fulfilled to realise an EP3 and hence provides direct physical insight into the design of high order EPs. For example, assuming a fixed host material (here taken as water with $\epsilon_h=1.76$), and interface seperation $d_1$ the constraints $\Delta = \Delta_1 = 0$ allow us to select appropriate dielectric materials (and corresponding required levels of gain) and nanowire radii, such that Eq.~\eqref{eq:EP3constraints} dictates the necessary metal permittivity to generate a third order degenerate point. Using known metallic dispersion equations (taken here as the Drude-Sommerfeld model with  $\epsilon_\infty = 5$, $\omega_p = 8.9~\text{eV}$, $\Gamma = 18~\text{fs}$ corresponding to silver \cite{Yang 2015}), the degenerate eigenfrequency can then be found. Assuming, $d_1=0.84$, $x_0 = 0.1$ and $m=1$, we find from the repeated root constraints that $d_2 = 2$ (yielding $D_1 = 199$~nm, $D_2 = 18$~nm and $D_3 = 6$~nm for $g = 2\times 10^{-8}$). Eq.~\eqref{eq:EP3constraints} then yields $\epsilon_m = (-1.548  + 0.856i)$ corresponding to a third order degeneracy at $\omega_{EP} = (8.352  + 0.250i) \times 10^{14}$~Hz. We neglect non-local effects in the material permittivities for simplicity \cite{deAbaj02008, Savage2012}.

  To verify the degeneracy of the modes predicted using the TO approach and its exceptional nature, we analyze the properties of the nanowire system in the vicinity of the expected EP3. In Figure~\ref{fig:EP3}(a) we present a plot of the logarithm of the scattering cross-section calculated as a function of gain and frequency using the method described in Ref.~\cite{Aubry 2010}. Selected line profiles (denoted I--V) are also shown in Figure~\ref{fig:EP3}(b), in which evolution of individual spectral peaks with material gain is illustrated. At low gains (line profile I and II), three spectral peaks are evident, which merge into a single broad peak at the EP3 (line profile III) before splitting again as gain is increased further (line profile IV and V). Accompanying blue data markers in Figure~\ref{fig:EP3}(b), found via rigorous finite element COMSOL simulations, show good agreement with the theoretical TO based calculations. Specifically, the average relative error and its standard deviation across the frequency range shown are 5.5\% and 0.17\% respectively.  Solid coloured lines in Figure~\ref{fig:EP3}(a) represent the real part of the eigenfrequencies, i.e. $\mbox{Re}[\omega_{r}]$, calculated via numerical solution of Eq.~\eqref{eq:cubic}. In general three distinct modes are found, however at the predicted EP the real part of the eigenfrequencies clearly coalesce. The inset in the lower right of Figure~\ref{fig:EP3}(a) shows the corresponding coallesence of the imaginary part of the eigenfrequency and illustrates the significant changes in loss associated with the emergence of the EP \cite{Wang 2023}. Specifically, in the PT-like unbroken phase, the mode corresponding to the green line exhibits relatively low loss, whereas in the PT-like broken phase, the loss increases sharply resulting in relatively weak scattering. In contrast, the loss for the other modes (blue and brown lines), decreases with increasing gain in the PT-like broken phase. This strong difference in loss means that at higher gains, only the two low loss spectral peaks are visible in the scattering spectrum (line profiles IV and V).

  Degeneracy in the complex eigenfrequencies, whilst suggestive, does not unequivocally demonstrate that solutions to Eq.~\eqref{eq:cubic} correspond to an EP. To demonstrate that the found degenerate point is indeed an EP3, we also show that the mode distributions for each eigenfrequency are identical at the degenerate point. Our TO method can fortunately provide direct information about the potential distributions for each plasmonic mode, which is often beyond the reach of traditional Hamiltonian approaches without additional modelling. Specifically, through Gaussian elimination it is simple to show that the solutions to Eq.~\eqref{eq:ME_3nd_order} take the form (with our material assumptions)
  \begin{align}
\left[\begin{array}{c} E_{y}(x_0-d_2)\\
  		E_y(x_0)\\
  		E_y(x_0+d_1)\end{array}\right]  = \left[\begin{array}{c} 1\\
  		\epsilon_h^{-1}\mathcal{M}(d_2,\epsilon_h,\epsilon_m) \\
  		\frac{\epsilon_m\mathcal{M}(d_2,\epsilon_h,\epsilon_m)}{\epsilon_h\mathcal{M}(d_2,\epsilon_m,\epsilon_d) }
  	\end{array}\right],
  \end{align}
  from which the full potential distributions follow. Example distributions are shown in Figure~\ref{fig:EP3}(c) for the points \Circled{1}-\Circled{9} on the resonance branches indicated in Figure~\ref{fig:EP3}(a). Color scales for each diagram span the respective maximum and minimum potential and are not necessarily the same for each panel. For non-degenerate conditions (\Circled{1}-\Circled{3} and \Circled{7}-\Circled{9})
  the nanowire structure exhibits standard plasmonic
  modes, including symmetric and asymmetric distributions at different boundaries. In contrast, at the degenerate point, the modes found for each branch are identical (\Circled{4}-\Circled{6}) as expected for an EP. Moreover, the distributions are highly confined   and exhibit unique spatial distributions not observed in lower order EPs. These patterns thus highlight the complexity and richness of the physical phenomena occurring at higher-order EPs.

 \begin{figure*}[t!]
 	\begin{center}
 		\includegraphics[width=0.75\textwidth]{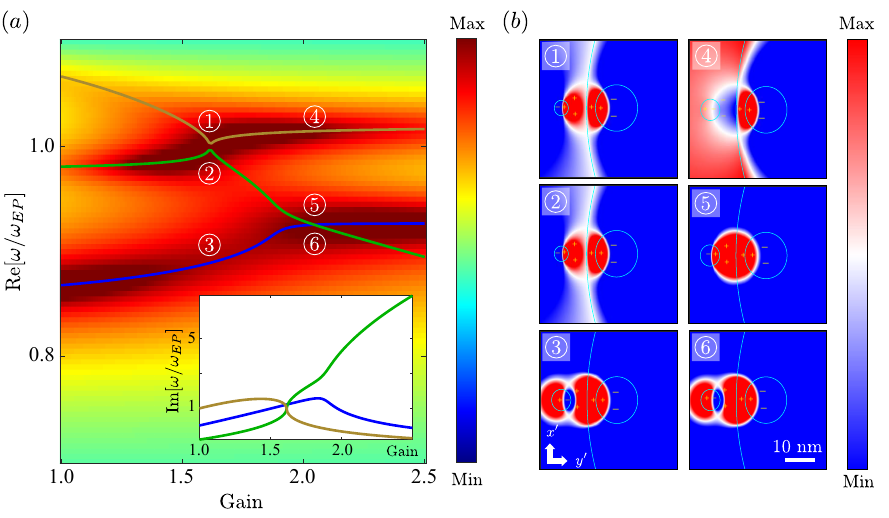}
 		\caption{\textbf{EP2 in core-shell/monomer coupled nanowire system} (a) As Figure~\ref{fig:EP3}(a) albeit for system parameters lying on the EA away from the EX (see Figure~\ref{fig:schematic}) (b) Potential distributions for \Circled{1}-\Circled{6} in (a), in which $+$ and $-$ denote a positive and negative potential respectively. Panels \Circled{5} and \Circled{6} correspond to the crossing of the real part of the eigenfrequecies, whereas the identical distributions in \Circled{1} and \Circled{2} signify an EP2. }
 		\label{fig:DP_EP_example}
 	\end{center}
 \end{figure*}

  Relaxing our previous constraint that $\Delta_1 = 0$, whilst still requiring that the discriminant of Eq.~\eqref{eq:cubic} be zero ($\Delta = 0$), implies that Eq.~\eqref{eq:cubic} possesses only two repeated roots \cite{cubicquartic}. We find that $\Delta = 0$ therefore defines curves in parameter space corresponding to systems supporting second (lower) order EPs. To illustrate this point we introduce perturbations $\delta\epsilon_h$ and $\delta\epsilon_d$, corresponding to variations in the complex electric permittivity of the host and dielectric gain material respectively, away from the EP3.  Figure~\ref{fig:schematic}(b) shows the resulting EP2 exceptional arcs (EA) \cite{Shen 2018, Zhou 2018} in the $(d_2, \delta\epsilon_h)$ domain when $\delta\epsilon_d=0$. Similarly, Figure~\ref{fig:schematic}(c) illustrates the case for perturbations in the gain material ($\delta\epsilon_d \neq 0$, $\delta\epsilon_h = 0$). Note that in each case two arcs result since $\Delta$ is quadratic in the corresponding system parameters. Through our TO analysis we can therefore easily determine the family of coupled nanowire configurations supporting EP2s, and how to tune particular parameters (such as wire radius) to move along these EAs to reach the point of convergence, i.e. the exceptional nexus (EX), and generate the resulting higher order EP3. We note that this insight and practical tunability is typically absent from Hermitian based design philosophies which consider more abstracted perturbations, e.g., to modal coupling constants, in contrast to the real physically relevant perturbations considered in our model.  Figure~\ref{fig:DP_EP_example}(a) analyzes the EP2 at a position along the exceptional EA, in a manner similar to Figure~\ref{fig:EP3}. We first note that, as shown in Figure~\ref{fig:DP_EP_example}(a), the eigenfrequencies corresponding to the brown and green solid lines possess the same real and imaginary parts for gains $\sim 1.6$, indicating the presence of an EP2, which is confirmed by the degenerate mode distributions shown in panels \Circled{5} and \Circled{6} of Figure~\ref{fig:DP_EP_example}(b). It is interesting, to note that at a gain of $\sim 2$, the real parts of the eigenfrequencies traced by the blue and green curves are equal, however their imaginary parts (inset of Figure~\ref{fig:DP_EP_example}(a)) remain different demonstrating this point is not a true degeneracy \cite{Chen 2020}.

\begin{figure}[t!]
	\begin{center}
		\includegraphics[width=\columnwidth]{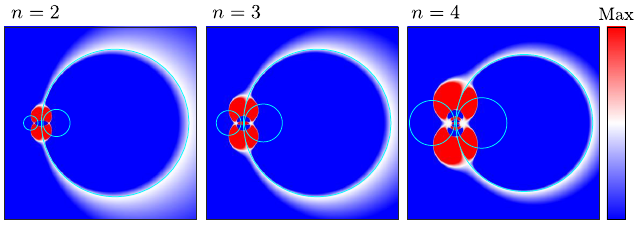}
		\caption{\textbf{EP3s with different angular momenta.} Potential distributions for modes with angular momenta (left) $n=2$, (middle) $n=3$ and (right) $n=4$, evaluated at an EP3 in a system analagous to that shown in Figure~\ref{fig:EP3}.}
		\label{fig:angular_momenta}
	\end{center}
\end{figure}

To further demonstrate the capabilities and versatility of our method in designing higher-order EPs, we have explored two further scenerios. Firstly, we consider design of EPs with different angular momenta, $n$. The corresponding potential distributions found for third order EPs in a core-shell/monomer nanowire system with angular momenta of $n=2$, $n=3$ and $n=4$ are shown in Figure~\ref{fig:angular_momenta}.

\begin{figure*}[t]
	\begin{center}
		\includegraphics[width=\textwidth]{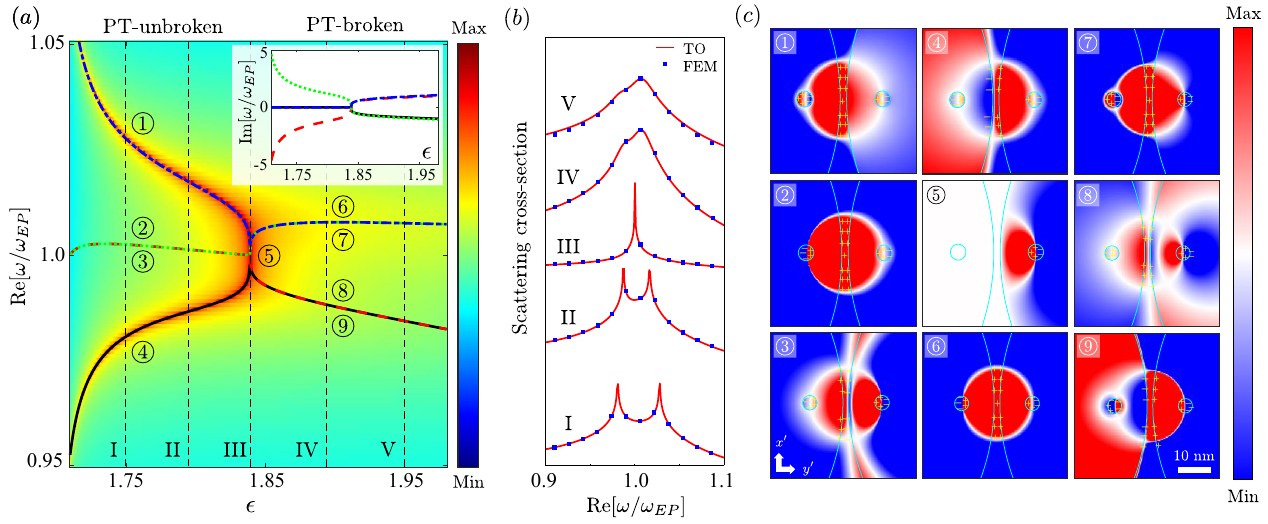}
		\caption{\textbf{EP4 in core-shell dimer nanowire system.} (a) Color density plot of the logarithm of the scattering cross-section of the coupled core-shell dimer structure as a function of $\epsilon$ and frequency. Solid and dashed lines represent the real parts of the four eigenfrequencies (corresponding imaginary parts are shown in the inset). (b) Scattering cross-sections corresponding to I-V in (a) as found using our TO approach (red lines) and finite element simulations (blue markers). (c) Potential distributions for eigenmodes at points labelled \Circled{1}-\Circled{9} in (a). The $+$ and $-$ markers denote a positive and negative potential respectively. Cyan lines demark the nanowire interfaces (only a small part of core-shell nanowires are shown to enhance visibility). Panel \Circled{5} corresponds to the EP3. }
		\label{fig5:EP4}
	\end{center}
\end{figure*}

Secondly, we consider design of a fourth-order exceptional point (EP4). Specifically, by considering a 5 layer gain-loss multilayer structure distributed symmetrically around the $y$-axis, we can generate, through application of Eq.~\eqref{eq:conformal_transform}, a PT symmetric configuration of two adjacent core-shell structured nanowires. The metallic outer shell of one nanowire is assumed to be described by electric permittivity $\epsilon_m = \epsilon_\infty - \omega_p^2 / \omega^2 + i \alpha$, whereas the inner dieletric core has $\epsilon_d = \epsilon - i \beta$, where $\alpha$ and $\beta$ are frequency independent parameters, $\epsilon = 1.84$, $\epsilon_\infty = 5$ and $\omega_p = 8.9$~eV. PT symmetry therefore implies for the other nanowire the core and shell have permittivities $\epsilon_m^*$ and $\epsilon_d^*$ respectively. The host material is taken to be air with unity permittivity. Figure~\ref{fig5:EP4}(a) shows the resulting distribution of the logarithmic scattering cross-section as a function of $\epsilon$ and frequency.  Using the assumed analytic forms of the permittivity functions, the resonance condition can be expressed as a polynomial in frequency $\omega$, specifically, $A\omega^8 + B\omega^6 + C\omega^4 + D\omega^2 + E = 0$, which is quartic in $\omega^2$ with three distinct roots. As before, the curves in Figure~\ref{fig5:EP4}(a) represent the real part of the eigenfrequencies, $\mbox{Re}[\omega_{r}]$, found from the corresponding resonance condition (see Supplementary Material). To aid visibility of overlapping branches, some curves are plotted using a dashed line style. The inset to Figure~\ref{fig5:EP4}(a) shows the corresponding imaginary part of the eigenfrequencies as a function of relative permittivity, which exhibit significantly different behaviour between the PT-unbroken and broken phases. Symmetry in the core-shell structures also leads to the existence of dark modes, resulting in weak scattering (and low scattering cross-sections) in spite of small modal losses \cite{Gao2018,Gomez2013}. The mode distributions can again be found (see Supplementary Material) and are shown at assorted positions along the resonance branches in Figure~\ref{fig5:EP4}(c) as marked \Circled{1}-\Circled{9} in Figure~\ref{fig5:EP4}(a). Constraints for the existence of multiple repeated roots (2, 3, or 4) are known \cite{cubicquartic} and can therefore be used to design exceptional points of higher order. In our case we seek an EP4 and use the constraints $8AC-3B^2 = B^3 -4ABC+8A^2D=16AB^2C-64A^2BD-3B^4 + 256A^3E=0$, which yields an EP4 at position \Circled{5} in parameter space. The good agreement between TO based predictions and rigorous COMSOL simulation results for the scattering cross-section, shown in Figure~\ref{fig5:EP4}, validates our approach. Specifically, the absolute relative error was calculated to have a mean of $\sim 0.54$\% and a standard deviation of $\sim 0.16$\%. Agreement between analytic and simulated mode distributions was also found.

\section{Conclusion}
This work has introduced a novel TO based approach for designing high-order EPs in nanoplasmonic systems. By utilising TO, we have mapped solutions from simple multilayer geometries to complex nanowire structures, enabling the design of EPs with targeted properties. Specifically, we have demonstrated the realization of an EP3 in a coupled core-shell/monomer nanowire system, identifying the precise constraints on system parameters required to achieve the desired degeneracy. Finite element simulations were used to confirm the existence and exceptional nature of the EP3.  Furthermore, we have shown how this approach can be extended to design EP2s along exceptional arcs, revealing a pathway to higher-order EPs through parameter space manipulation.  The versatility of our method was further highlighted by the demonstration of EP3s with varying angular momenta and the design of a EP4 in a coupled core-shell dimer system.  Our TO-based methodology overcomes limitations of traditional Hamiltonian approaches, which often struggle to connect abstract model parameters to experimentally controllable variables. Moreover, our approach naturally yields corresponding mode distributions without requiring additional modelling. Accordingly, the TO methodology offers a powerful and flexible framework for engineering EPs in complex non-Hermitian nanophotonic structures. {Experimentally, introducing moderate optical gain adjacent to metals, e.g. through use of dyes, fluorescent polymers or quantum dots, has been shown to enable compensation of plasmonic losses and nanoscale lasing \cite{DeLeon2010,Oulton2009,Gather2010,Kress2017}, thereby supporting the practicality of the gain-assisted exceptional points considered here. Our work therefore} opens new avenues for applications in sensing, light manipulation, and other areas where precise control over light-matter interaction is crucial.

\medskip

% Acknowledgements
\medskip
\textbf{Acknowledgements} \par

K.W. is funded by an Interdisciplinary Graduate Programme PhD Research Scholarship through the Institute for Digital Molecular Analytics and Science (IDMxS) under the Singapore Ministry of Education Research Centres of Excellence scheme (EDUN C-33-18-279-V12). MRF acknowledges further funding from Singapore Ministry of Education Academic Research Fund (Tier 1) Grant No. RS13/23. Y.L. acknowledges funding support from Fundamental Research Funds for the Central Universities, NUAA(No. NE2024007).

% References
\medskip

% Use the following code if you wish to generate your bibliography with BibTeX;
% replace the string "MSP-template" below with the name(s) of
% the BibTeX data base(s) you want to use.
% The resulting bibliography-output (the content of the .bbl file)
% must be pasted back into this file before submission.
% Please also include your BibTeX data base file(s) in your submission
% so that we can re-run BibTeX if necessary.
%
%\bibliographystyle{MSP}
%\bibliography{MSP-template}

\textbf{References}\\

\end{document}